\documentclass[epj]{svjour}

\usepackage{graphicx,amssymb}
\usepackage[numbers,sort&compress,merge]{natbib}

\begin{document}
\pagestyle{plain}
\title{Light Neutralino Dark Matter in the pMSSM}
\subtitle{Implications of LEP, LHC and Dark Matter Searches on SUSY Particle Spectra}
\author{A. Arbey\inst{1}\inst{2}\inst{3} \and 
        M. Battaglia\inst{3} \inst{4} \inst{5} \and
        F. Mahmoudi\inst{3} \inst{6}
}                     

\institute{Centre de Recherche Astrophysique de Lyon, Observatoire de Lyon, Saint-Genis Laval Cedex, F-69561, France; CNRS, UMR
5574; Ecole Normale Sup\'erieure de Lyon, Lyon, France
\and Universit\'e de Lyon, France; Universit\'e Lyon 1, 
F-69622~Villeurbanne Cedex, France \and
CERN, CH-1211 Geneva 23, Switzerland \and
Santa Cruz Institute of Particle Physics, University of California, Santa Cruz,
CA 95064, USA \and
Lawrence Berkeley National Laboratory, Berkeley, CA 94720, USA \and
Clermont Universit\'e, Universit\'e Blaise Pascal, CNRS/IN2P3, LPC, BP 10448, F-63000 Clermont-Ferrand, France}
\date{}

\abstract{
We investigate the viability of light neutralino scenarios as promoted by dark matter direct detection experiments. Using high statistics scans in the pMSSM we have identified several scenarios which give rise to very light neutralinos with large direct detection scattering cross sections.
Our results are challenged with constraints from dark matter relic density, direct detection, indirect detection, as well as flavour physics, electroweak precision tests, LEP and Tevatron limits, LHC limits on SUSY, Higgs and monojet searches. In particular we study the effect of a Higgs boson in the range 122.5 $< M_{h} <$ 127.5 GeV. We show that several scenarios emerge in agreement with all the constraints, and we study their characteristics and the LHC sensitivity.
} 
\maketitle
\section{Introduction}
\label{sec:1}

One of the appealing features of Supersymmetry (SUSY), which has contributed to promote it to 
the role of template  model of new physics beyond the Standard Model, is the natural inclusion of a 
weakly interacting massive particle (WIMP) as cold dark matter (DM) candidate. In most of the cases, 
this is the lightest neutralino, $\tilde \chi^0_1$, which is the lightest supersymmetric particle (LSP).  
Scenarios with neutralino LSP realise the exact amount of dark matter relic density to match 
the precise determination obtained in the study of cosmic microwave background (CMB). This happens 
for neutralino masses of ${\cal{O}}$(100~GeV) due to the suggestive fact that a particle with 
$\sim$100~GeV mass and typical weak interaction couplings gives the correct neutralino relic density, 
$\Omega_{\mathrm{CDM}} h^2 \sim$0.1, in standard cosmology models which often referred to as the ``WIMP miracle''.

The analysis of constraints from low energy data, dark matter density and the results of the 
LHC searches in generic SUSY scenarios shows that light neutralinos are still well compatible with 
current data. While searches for signals of strongly interacting SUSY particles in $pp$ collisions using 
missing transverse energy (MET) signatures are ongoing at the LHC, ground-based direct detection DM experiments, 
such as XENON~\cite{Aprile:2011hi} and CDMS~\cite{Ahmed:2009zw}, have reached a sensitivity in spin-independent 
WIMP-nucleon cross section at WIMP masses of ${\cal{O}}$(100~GeV), which covers a significant fraction of the 
SUSY parameter space explored at the LHC. 
The present XENON-100 limit~\cite{Aprile:2011hi} covers $\sim$20\% of the MSSM parameter space compatible with 
low energy data and the relic dark matter constraint~\cite{Arbey:2011aa}.

The possibility that the LSP mass is much smaller than the electroweak scale, is put forward by three dark matter 
direct detection experiments, which have reported a possible signal of WIMP interaction corresponding to very light 
particles, 5 $< M_{\mathrm{CDM}} <$ 15~GeV close to the edge of the XENON-100 and CDMS sensitivity. 
These are the DAMA experiment~\cite{Bernabei:2000qi,Bernabei:2010mq} at the Laboratory Nazionali del 
Gran Sasso, Italy, the CoGENT experiment~\cite{Aalseth:2010vx,Aalseth:2011wp} in the Soudan mine and 
the CRESST experiment~\cite{Angloher:2011uu}, also at Gran Sasso. 
While there is still substantial debate on the interpretation of these data and the compatibility of 
the reported results with the exclusion bounds established by the CDMS and XENON 
experiments~\cite{Savage:2008er,Fitzpatrick:2010em,Kelso:2011gd,Gelmini:2012dq} 
an agreement could only be possible in the low neutralino mass region. Therefore it is interesting to 
explore the feasibility of light neutralino solutions in generic SUSY models and their compatibility with the 
results of the LEP and LHC searches. 

In this study we consider SUSY scenarios with the lightest neutralino mass in the range 5 $< M_{\chi^0_1} <$ 40~GeV 
in the 19-parameter pMSSM model. The pMSSM is a minimal supersymmetric extension of the Standard Model (MSSM) 
with $R$-parity conservation, where the masses of all SUSY particles are treated as free parameters \cite{Djouadi:1998di}. 
The use of the pMSSM enables us to access scenarios which are not available in constrained SUSY models, such as the CMSSM. 

There have been already several studies of the SUSY parameter space with light 
neutralinos~\cite{Hooper:2002nq,Bottino:2002ry,Bottino:2003cz,Profumo:2008yg,Bottino:2008xc,Dudas:2008eq,Feldman:2010ke,Kuflik:2010ah,Belikov:2010yi,Vasquez:2010ru,Choi:2011vv,AlbornozVasquez:2011yq,Bhattacharyya:2011se,Cumberbatch:2011jp,Calibbi:2011ug,Bottino:2011xv,Calibbi:2011un}. 
They considered either a constrained MSSM scenario or more generic analyses in effective MSSM scenarios where gaugino masses 
are not unified at the GUT scale. In addition, analyses within the NMSSM scenarios were also performed~\cite{Barger:2005hb,Gunion:2005rw,Das:2010ww,Gunion:2010dy,Gunion:2010dy,Cumberbatch:2011jp,Carena:2011jy}. 
Compared to those studies we consider here a broader phase space of parameters and study for the first time the light neutralinos in the 
pMSSM in its full glory with 19 free parameters. In particular, we do not assume degenerate masses for the right and left handed squarks, 
contrary to the previous works. By this choice, different squark mixings are allowed and scenarios with light squarks and reduced couplings 
to the $Z^0$ boson can be realised in our scans of the parameter space. This study reveals therefore scenarios not yet identified in the earlier works. 
We employ a large statistics of more than 500~M pMSSM points for a generic scan and more than 500~M extra points for specific scans, and impose the 
latest constraints from the LHC data, including those from Higgs and monojet searches. In particular we study the effect of a Higgs boson in the range 122.5 $< M_{h} <$ 127.5 GeV. 
We use a realistic simulation, validated on the results of full simulation and reconstruction to study the 
phenomenology of these scenarios and the response of the standard SUSY searches on 5~fb$^{-1}$ of data 
at 7~TeV~\cite{Arbey:2011un}.

This paper is organised as follows. Section~2 describes the region highlighted by the results of the 
DAMA, CoGENT and CRESST direct dark matter searches and the pMSSM simulation with the constraints applied
to reproduce bounds from flavour physics, electroweak data, earlier SUSY searches at LEP-2 and the Tevatron
and dark matter relic density. The different viable pMSSM scenarios with a light neutralino are presented.
Section~3 discusses  their phenomenology at the LHC and the impact of the SUSY searches based on jets, leptons 
and missing transverse energy. Section~4 has the conclusions.

\section{The pMSSM Phase Space with light neutralino LSP}
\label{sec:2}

\subsection{Simulation and Tools}
\label{sec:2-1}

The study of the pMSSM parameter space with light neutralinos is based on the combination of several programs from 
spectrum generation to the computation of dark matter scattering cross sections and relic density, flavour and 
electroweak observables, as well as the simulation and analysis of events in 7~TeV $pp$ collisions. This study 
is part of a broad program on the implications of LHC results for the MSSM through scans of the pMSSM. A detailed 
description of the software tools employed is given in~\cite{Arbey:2011un}. We mention here only the software tools 
most relevant to this study. SUSY spectra are generated with {\tt SOFTSUSY 3.2.3}~\cite{Allanach:2001kg}. The widths 
and decay branching fractions of SUSY particles are computed using {\tt SDECAY 1.3} \cite{Muhlleitner:2003vg}. 
The dark matter relic density is calculated with {\tt SuperIso Relic v3.2}~\cite{Mahmoudi:2007vz,*Mahmoudi:2008tp,Arbey:2009gu}, 
which provides us also with the flavour observables. {\tt MicrOMEGAs 2.4} \cite{Belanger:2008sj} is used to compute 
neutralino-nucleon scattering cross-sections. Corrections to the $Z^0$ electroweak observables are calculated 
analytically. We perform flat scans of the 19 pMSSM parameters within the ranges summarised in Table~\ref{tab:paramSUSY}.
For this study, over one billion pMSSM points have been generated in total.
\begin{table}
\begin{center}
\begin{tabular}{|c|c|}
\hline
~~~~Parameter~~~~ & ~~~~~~~~Range~~~~~~~~ \\
\hline\hline
$\tan\beta$ & [1, 60]\\
\hline
$M_A$ & [50, 2000] \\
\hline
$M_1$ & [-300, 300] \\
\hline
$M_2$ & [-650, 650] \\
\hline
$M_3$ & [0, 2500] \\
\hline
$A_d=A_s=A_b$ & [-10000, 10000] \\
\hline
$A_u=A_c=A_t$ & [-10000, 10000] \\
\hline
$A_e=A_\mu=A_\tau$ & [-10000, 10000] \\
\hline
$\mu$ & [-3000, 3000] \\
\hline
$M_{\tilde{e}_L}=M_{\tilde{\mu}_L}$ & [0, 2500] \\
\hline
$M_{\tilde{e}_R}=M_{\tilde{\mu}_R}$ & [0, 2500] \\
\hline
$M_{\tilde{\tau}_L}$ & [0, 2500] \\
\hline
$M_{\tilde{\tau}_R}$ & [0, 2500] \\
\hline
$M_{\tilde{q}_{1L}}=M_{\tilde{q}_{2L}}$ & [0, 2500] \\
\hline
$M_{\tilde{q}_{3L}}$ & [0, 2500] \\
\hline
$M_{\tilde{u}_R}=M_{\tilde{c}_R}$ & [0, 2500] \\
\hline
$M_{\tilde{t}_R}$ & [0, 2500] \\
\hline
$M_{\tilde{d}_R}=M_{\tilde{s}_R}$ & [0, 2500] \\
\hline
$M_{\tilde{b}_R}$ & [0, 2500] \\
\hline
\end{tabular}
 \end{center}
\caption{pMSSM parameter ranges adopted in the scans (in GeV when applicable).\label{tab:paramSUSY}}
\end{table}

Event generation of inclusive SUSY samples in  $e^+e^-$ and $pp$ collisions is performed with 
{\tt PYTHIA 6.4.24}~\cite{Sjostrand:2006za}. Cross sections for $pp$ collisions are rescaled to their NLO 
values by the k-factors obtained with {\tt Prospino 2.0}~\cite{Beenakker:1996ed}. Samples of $e^+e^-$ events 
are reconstructed at generator level to assess their observability at LEP-2. Hadronic jets are clustered using 
the PYCLUS algorithm with $d_{join} = $ 2.5~GeV. LHC generated events are passed through fast detector simulation 
using {\tt Delphes 1.9}~\cite{Ovyn:2009tx} tuned for the CMS detector. The event reconstruction follows the 
procedure of the CMS SUSY analyses as discussed in~\cite{Arbey:2011un}.

\subsection{Constraints}
\label{sec:2-2}

To constrain the pMSSM parameter space, we apply different limits, from cosmological data, flavour physics, electroweak data, and collider searches.

\subsubsection{Dark Matter}
\label{sec:2-2-1}

Dark matter constraints arise from the relic density determination, $\Omega_{\mathrm{CDM}} h^2$, 
mostly from the analysis of the WMAP data~\cite{Komatsu:2010fb}, and from direct detection experiments. 
For the relic density, we consider two intervals: the tight WMAP bound of 0.068$ < \Omega_{\chi} h^2 <$ 0.155 which includes the theoretical uncertainties. We study also the effects of a loose limit where 
we request that the neutralino contribution, $\Omega_{\chi} h^2$, is non zero and 
below, or equal to, the upper limit set by the WMAP result: 10$^{-4} < \Omega_{\chi} h^2 <$ 0.155. This 
leaves the possibility of other sources of dark matter alongside the lightest neutralino. In 
section~\ref{sec:2-4}, where we discuss non-standard scenarios, we further relax this requirement.

The results of dark matter direct detection experiments are more problematic. The claims from the experiments 
reporting an excess of events compatible with the signal of a light WIMP must be confronted with the 
limits obtained by XENON~\cite{Aprile:2011hi} and CDMS~\cite{Ahmed:2009zw}, which has also performed analyses 
relaxing the energy cut-off~\cite{Ahmed:2010wy} and, more recently, searched for an annual modulation of the event rate~\cite{Ahmed:2012vq}. In this study we accept pMSSM with 5 $\le M_{\chi^0_1} <$ 40~GeV and 10$^{-7} < \sigma_{\chi p}^{\mathrm{SI}} < $ 10$^{-3}$ pb, in order to be in the region where data could be reconciled.

Other cosmological bounds can come from indirect detection signatures, but they are subject to large cosmological and 
astrophysical uncertainties. We do not impose them as a constraints in the analysis but we comment on the consequences of the Fermi-LAT results \cite{Ackermann:2011wa} in 
Section~\ref{sec:2-3}.

\subsubsection{Flavour Physics}
\label{sec:2-2-2}

Flavour physics sets important constraints on the SUSY parameters. We impose bounds from $b$ and $c$ meson decays, 
which have been discussed in details in~\cite{Arbey:2011un}. In particular, the decay $B^0_s \to \mu^+ \mu^-$, which 
can receive extremely large SUSY contributions at large $\tan\beta$, deserves special attention~\cite{Akeroyd:2011kd}. 
An excess of events in this channels was reported by the CDF-II collaboration at the Tevatron~\cite{Aaltonen:2011fi} 
and upper limits have been set by the LHCb~\cite{Aaij:2012ac} and CMS~\cite{Chatrchyan:2012rg} collaborations at LHC. 
Recently the LHCb collaboration has presented their latest result for the search of this decay based on 1~fb$^{-1}$ 
of data. A 95\% C.L. upper limit on its branching fraction is set at $4.5 \times 10^{-9}$~\cite{Aaij:2012ac}, which 
closely approaches the SM prediction of (3.53 $\pm$ 0.38) $\times 10^{-9}$ for the rate of this process \cite{Mahmoudi:2012zz}. After accounting for theoretical uncertainties, estimated at the 10\% level, the constraint 
BR($B^0_s \to \mu^+ \mu^-$) $< 5 \times 10^{-9}$ is used in this analysis.

\subsubsection{Electroweak data}
\label{sec:2-2-3}

The precision $Z^0$ line-shape and other electroweak observables place serious constraints on new particles. In particular, the 
accurate measurements of the $Z^0$ total width and its partial  decay widths obtained at LEP provide a tight bound to 
the contribution from new particles with mass below $M_{Z^0}$/2.
The scenarios considered here with light $\chi^0_1$ are constrained from the neutralino 
contribution to the $Z^0$ invisible width.  We compute the SM $Z^0$ total width and that into neutrinos 
using {\tt ZFitter 6.42}~\cite{Bardin:1999yd} for the input parameter ranges 115 $< M_H <$ 145~GeV,  
$M_{\mathrm{top}}$ = (172.9$\pm$1.1)~GeV and $\alpha_s(M_Z^2)$ = 0.1184$\pm$0.0007. We obtain a SM $Z^0$ total width  
$\Gamma_{\mathrm{tot}}$ = (2494.83 $\pm$ 0.54)~MeV and invisible width $\Gamma_{\mathrm{inv}}$ = (501.62$\pm$0.10)~MeV,
where the uncertainty reflects the range of values used for the input parameters. These have to be compared 
to the average of LEP measurements giving (2495.2$\pm$2.3) and (499.0$\pm$1.5) MeV, respectively. 
We excludes light neutralinos with a contribution to the $Z^0$ invisible width larger than 3~MeV. This restricts the acceptable 
points to those where the $\tilde \chi^0_1$ is bino-like and its contribution to the $Z^0$ width, $\Gamma_{\chi}$, is negligible 
so it can evade the LEP electroweak bounds and corresponds to relatively large values of the higgsino mass parameter $|\mu|$.

The $Z^0$ total width constrains the masses of the lightest chargino, $\tilde \chi^{\pm}_1$, and of the squarks of the 
first two generations, $\tilde q$, to be above 45~GeV. Squarks of the third generation can evade these constraints for 
specific values of their $\theta_{\tilde q}$ angle in the mass mixing matrix, corresponding to a vanishing $Z^0$ 
coupling, $I_3^q \cos^2 \theta_{\tilde q} - Q_{\tilde q} \sin^2 \theta_W$. We also compute the 
$\tilde{b_1} \bar{\tilde b}_1$ contribution to the $Z^0$ width for each generated pMSSM points and require it to be 
smaller than 5~MeV, to satisfy the hadronic $Z^0$ width limits.

\subsubsection{SUSY searches at LEP-2 and Tevatron}
\label{sec:2-2-4}

The general constraints on SUSY particle masses from direct searches at lower energy 
colliders are summarised in Table~\ref{tab:constraintsSUSY}.
\begin{table}
\begin{center}
\begin{tabular}{|c|c|c|}
\hline
Particle & Limits & ~~~~~~Conditions~~~~~~\\
\hline\hline
$\tilde \chi^0_2$ & 62.4 & $\tan\beta < 40$\\
\hline
$\tilde \chi^0_3$ & 99.9 & $\tan\beta < 40$\\
\hline
$\tilde \chi^0_4$ & 116 & $\tan\beta < 40$\\
\hline
$\tilde \chi^\pm_1$ & 92.4 & $m_{\tilde \chi^\pm_1} - m_{\tilde \chi^0_1} < 4$ GeV\\
 & 103.5 & $m_{\tilde \chi^\pm_1} - m_{\tilde \chi^0_1} > 4$ GeV\\
\hline
$\tilde{e}_R$ & 73 & \\
\hline
$\tilde{e}_L$ & 107 & \\
\hline
$\tilde{\tau}_1$ & 81.9 & $m_{\tilde{\tau}_1} - m_{\tilde \chi^0_1} > 15$ GeV\\
\hline
$\tilde{u}_R$ & 100 & $m_{\tilde{u}_R} - m_{\tilde \chi^0_1} > 10$ GeV\\
\hline
$\tilde{u}_L$ & 100 & $m_{\tilde{u}_L} - m_{\tilde \chi^0_1} > 10$ GeV\\
\hline
$\tilde{t}_1$ & 95.7 & $m_{\tilde{t}_1} - m_{\tilde \chi^0_1} > 10$ GeV\\
\hline
$\tilde{d}_R$ & 100 & $m_{\tilde{d}_R} - m_{\tilde \chi^0_1} > 10$ GeV\\
\hline
$\tilde{d}_L$ & 100 & $m_{\tilde{d}_L} - m_{\tilde \chi^0_1} > 10$ GeV\\
\hline
& 248 & $m_{\tilde \chi^0_1} < 70$ GeV, $m_{\tilde{b}_1} - m_{\tilde \chi^0_1} > 30$ GeV\\
& 220 & $m_{\tilde \chi^0_1} < 80$ GeV, $m_{\tilde{b}_1} - m_{\tilde \chi^0_1} > 30$ GeV\\
$\tilde{b}_1$ & 210 & $m_{\tilde \chi^0_1} < 100$ GeV, $m_{\tilde{b}_1} - m_{\tilde \chi^0_1} > 30$ GeV\\
& 200 & $m_{\tilde \chi^0_1} < 105$ GeV, $m_{\tilde{b}_1} - m_{\tilde \chi^0_1} > 30$ GeV\\
& 100 & $m_{\tilde{b}_1} - m_{\tilde \chi^0_1} > 5$ GeV\\
\hline
$\tilde{g}$ & 195 & \\
\hline
\end{tabular}
\end{center}
\caption{Constraints on the SUSY particle masses (in GeV) from searches at LEP and the Tevatron 
\label{tab:constraintsSUSY}}
\end{table}
However, we note that constraints from SUSY searches at LEP-2 and Tevatron can be evaded in case of small 
mass splittings with the lightest neutralino, corresponding to low energy, or transverse energy, released 
in the detector or to vanishing couplings. Since these are crucial for assessing the viability of several of 
the small mass splitting solutions highlighted in this study, we explicitly study the detectability of the 
points fulfilling our selection criteria. For each point we generate an inclusive SUSY sample of 10k events in 
$e^+e^-$ collisions at $\sqrt{s}$ = 208~GeV and apply the LEP-2 reconstruction and selection criteria of the 
analyses of~\cite{Abdallah:2003xe}. 
First, we consider the $\chi^+_1 \chi^-_1$ and $\chi^0_2\,\chi^0_1$ channels and adopt the SM backgrounds 
estimated in the original analyses for these channels. We compute the cross sections and the number of signal 
events passing the selection criteria. We compare these cross sections to the minimum excluded value as a 
function of $M_{\chi^{\pm}_1}$ and $\Delta M = M_{\chi^{\pm}_1} - M_{\chi^{0}_1}$ from the final combination of 
the LEP-2 results. We reject points which give a cross section in excess to that excluded by the combined LEP-2 
data for the same $\Delta M$ and a number of selected signal events larger than the SM background. A first estimate indicates that chargino masses smaller than 40 GeV are excluded by LEP data, independently from $\Delta M$. Therefore, we impose the constraint $M_{\chi^\pm_1} > 40$ GeV in the following, and we will analyse a posteriori the exclusion of the points passing all the other constraints.

Then, we consider the $\tilde b_1 \bar{\tilde b}_1$ channel for small $\Delta M = M_{\tilde b_1} - M_{\chi^{0}_1}$ values. 
We simulate $e^+e^- \to \tilde b_1 \bar{\tilde b}_1$ at $\sqrt{s}$ = 208~GeV for each point fulfilling our cuts. 
Again, we compare the production cross section to the minimum value excluded by the combination of the results 
of the LEP-2 experiments as a function of $\Delta M$ and reject points with cross sections in excess to this value.

\subsubsection{Higgs and SUSY searches at LHC}
\label{sec:2-2-5}

The searches conducted by ATLAS and CMS on the 7~TeV data have already provided a number of 
constraints relevant to this study. First, the MET analyses~\cite{Aad:2011ib,Chatrchyan:2011zy} 
have excluded a fraction of the 
MSSM phase space corresponding to gluinos below $\sim$600~GeV and scalar quarks of the first 
two generations below $\sim$400~GeV. These are included using the same analysis discussed 
in~\cite{Arbey:2011un}, extended to an integrated luminosity of 5~fb$^{-1}$. 
Then, the search for the $A^0 \to \tau^+ \tau^-$ decay~\cite{Aad:2011rv,Chatrchyan:2012vp} has excluded 
a significant fraction of the $M_A$ -- $\tan \beta$ plane at low values of $M_A$ and large to 
intermediate values of $\tan \beta$.

But the LHC 2011 data has not only brought exclusions. The excess of events on the $\gamma \gamma$, 
four-lepton and $\ell \nu \ell \nu$ final states reported by both ATLAS~\cite{ATLAS:2012ae} and 
CMS~\cite{Chatrchyan:2012tx} provide an intriguing hint for a Higgs boson. If these events are 
the first signal of a boson with mass $M_h \simeq$ 125~GeV, they will represent a significant 
constraints on our scenarios, in particular for their implication on the scalar top 
mass~\cite{Arbey:2011ab}. Here, we require 122.5 $< M_h <$ 127.5~GeV, which corresponds to the 
mass region allowed by the current data.

Finally, the results of the searches for monojets and isolated photons~\cite{Chatrchyan:2012aa} 
can be used to set constraints on the quark-neutralino couplings and thus on the $\chi p$ scattering 
cross-sections~\cite{Bai:2010hh}. In particular, the constraints derived for the spin-dependent cross-sections 
are more severe than those from direct detection experiments. Contrary to dark matter direct searches, 
the collider limits do not depend on astrophysical assumptions, such as the inferred local density of dark 
matter, which suffers from large uncertainties.

In this analysis we apply the constraints on both spin-dependent and spin-independent $\sigma_{\chi p}$ 
from the LHC searches to our pMSSM points.

\subsection{Allowed Regions}
\label{sec:2-3}

We select the valid pMSSM points fulfilling the requirements discussed above; in particular the light neutralino mass in the range 
5 $< M_{\tilde \chi^0_1} <$ 40~GeV, relic dark matter density as given in section \ref{sec:2-2-1} and 
spin-independent scattering cross-section $10^{-7} < \sigma_{\chi p}^{\mathrm{SI}} < 10^{-3}$~pb.

\begin{figure}[t!]
\begin{center}
\includegraphics[width=0.5\textwidth]{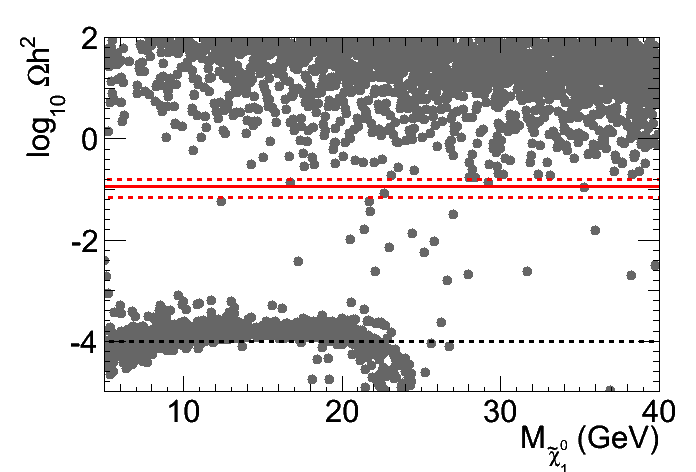}\\
\includegraphics[width=0.5\textwidth]{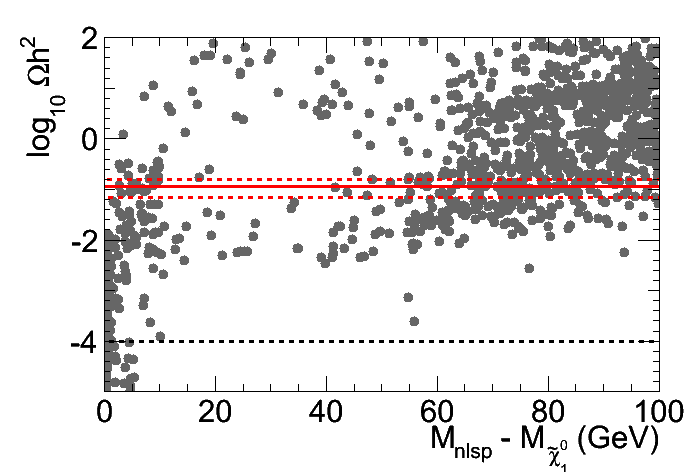}
\end{center}
\caption{Relic density in function of the $\tilde \chi^0_1$ mass (upper panel) and NLSP mass (lower panel). All the points in these plots pass the constraints described in section \ref{sec:2-2}. The continuous red line corresponds to the WMAP dark matter central values, and the dashed red lines to the upper and lower limits that we impose. The dashed black line corresponds to the lower limit that we use when relaxing the lower WMAP constraint.}
\label{fig:Oh2}
\end{figure}

\begin{figure}[t!]
\begin{center}
\includegraphics[width=0.5\textwidth]{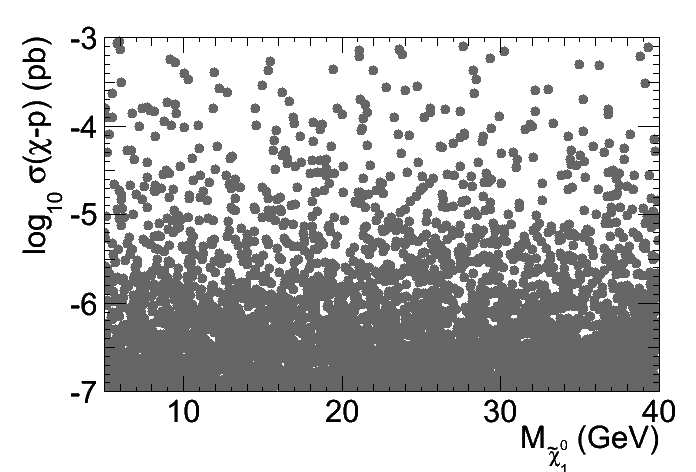}\\
\includegraphics[width=0.5\textwidth]{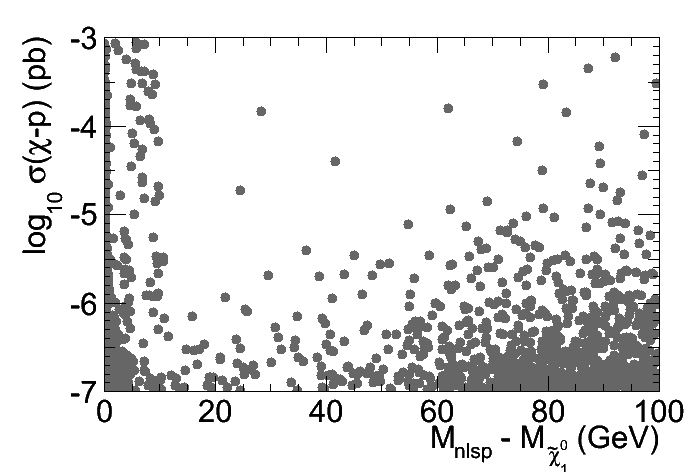}
\end{center}
\caption{Spin independent $\chi$-$p$ scattering cross-section as a function of the $\tilde \chi^0_1$ mass (upper panel) and NLSP mass 
(lower panel). All the points in these plots pass the constraints 
described in section \ref{sec:2-2}.}
\label{fig:sig}
\end{figure}

In general, for such a light LSP the relic density is larger than the upper relic density constraint, as shown in Fig.~\ref{fig:Oh2}. 
However, if the mass difference between the next lightest supersymmetric particle (NLSP) and the LSP, $\Delta M$, is small enough, 
the relic density is strongly decreased by the coannihilations of the two particles in the early universe and $\Omega_{\chi} h^2$ 
can even fall below the lower limit applied for this study. pMSSM points compatible with the relic density constraint have values of the 
$\Delta M$ mass splitting of just a few GeV.
The situation for the scattering cross-section is reversed: most of the selected pMSSM points have small scattering cross-sections, 
as shown in Fig.~\ref{fig:sig}. In order to increase it up to the values highlighted by the direct search experiments claiming a 
light WIMP signal, $\sigma_{\chi p}^{\mathrm{SI}} \sim 10^{-6}$~pb, we have to require $\Delta M$ values below 1~GeV, which corresponds 
to a requirements opposite to that found for the relic density. Therefore, we can respect all the constraints imposed on the pMSSM points 
with light neutralinos only for few, very specific scenarios.

\begin{figure}[t!]
\begin{center}
\hspace*{0.8cm}\includegraphics[width=0.2\textwidth]{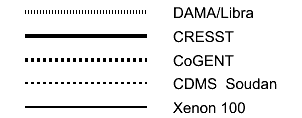}\\
\includegraphics[width=0.5\textwidth]{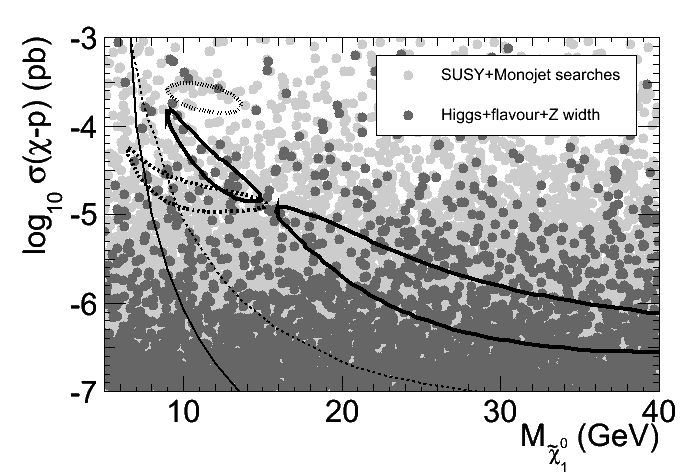}
\end{center}
\caption{Spin independent $\chi$-$p$ scattering cross-section as a function of the $\tilde \chi^0_1$ mass. The light gray points are consistent with the 
direct and monojet search limits (Higgs searches excluded). The dark gray 
points pass the constraints from flavour physics, the $Z$ decay constraints, and the Higgs mass limit $122.5 < M_h < 127.5$~GeV. The thick 
contour lines correspond to the zones favoured by DAMA, CRESST and CoGENT, and the thin lines to the exclusion limits by CDMS and 
XENON.}\label{fig:sig_norelic}
\end{figure}

\begin{figure}[t!]
\begin{center}
\includegraphics[width=0.5\textwidth]{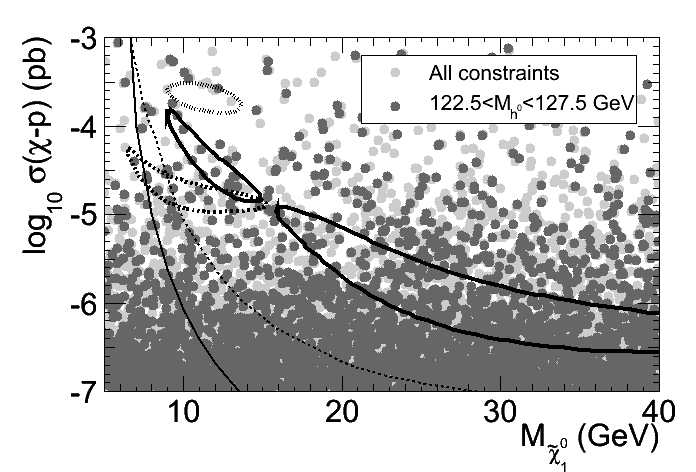}
\end{center}
\caption{Spin independent $\chi$-$p$ scattering cross-section as a function of the $\tilde \chi^0_1$ mass. The light gray colour corresponds to 
the points passing all the constraints presented in Fig.~\ref{fig:sig_norelic}, except the Higgs mass limit $122.5 < M_h < 127.5$~GeV. The dark gray points also pass the Higgs mass limits. The contour lines are the same as in Fig.~\ref{fig:sig_norelic}.}\label{fig:sig_higgs}
\end{figure}

The effect of the collider data, SUSY and monojet searches, flavour and electroweak constraints, on the pMSSM points generated with our scan is shown
in Fig.~\ref{fig:sig_norelic}. The collider constraints decrease the number of points without significantly modifying their distribution in the 
$M_{\tilde \chi^0_1}$ -- $\sigma^{SI}_{\chi p}$ space. However, the electroweak constraints have a significant impact in removing points at large 
scattering cross-sections, especially for larger values of the neutralino mass.
The constraints from Higgs searches alone is shown in Fig.~\ref{fig:sig_higgs}. Again, the Higgs mass constraint reduces the number of points, but does 
not modify their overall distribution.
Using the points from the generic pMSSM scan, we find 20 and 5 points passing all the selections in the region of interest, for the loose and tight 
WMAP constraints, respectively. The effect of the various selection criteria on the number of scan points retained 
is summarised in Table~\ref{tab:statgen}. The relic density constraint in this region of the parameter space is rejecting a particularly 
large fraction of pMSSM points. This underscores the difficulty to harmonise the large scattering cross-section corresponding to the possible light WIMP 
signals and the WMAP results discussed above.
\begin{table}
\begin{center}
\begin{tabular}{|l|c|c|c|}
\hline
Selection & pMSSM &  Selection &  Cumulative \\
& points &  Efficiency &  Efficiency \\
\hline
\hline
Valid points with & 1~M & -- & -- \\
light $\chi^0_1$, large $\sigma(\chi-p)$ & & & \\
\hline
Monojet searches & 280~k & 0.28 & 0.28 \\
\hline
SUSY searches & 90~k & 0.33 & 0.09 \\
\hline
LEP searches & 50~k & 0.60 & 0.05 \\
\hline
Flavour physics & 20~k & 0.37 & 0.02 \\
\hline
Higgs searches & 10~k & 0.47 & 0.01 \\
\hline
Loose WMAP limit& 20 & $2\times10^{-3}$ & $2\times10^{-5}$  \\
\hline
Tight WMAP limit& 5 & 0.25 & $5\times10^{-6}$  \\
\hline
\end{tabular}
\caption{Scan statistics for the generic scan.\label{tab:statgen}}
\end{center}
\end{table}
In order to improve the statistics for specific scenarios, we perform specific scans within restricted parameter sets starting from the points 
passing the relic density constraints. We identify three distinct classes of pMSSM solutions:
i) the NLSP is a slepton slightly above the LEP limit, with a neutralino of about 30 GeV
ii) the lightest chargino is degenerate with the  $\tilde \chi^0_1$, often with a compressed gaugino spectrum and light Higgs bosons and 
iii) a scalar quark is degenerate with the $\tilde \chi^0_1$ while other scalar quarks and leptons are relatively heavy. Given the Higgs 
mass constraints, the possible light squarks for class iii) are those of the first and second generations 
or the lightest scalar bottom quark, $\tilde b_1$.

\subsubsection{$\tilde \ell$ NLSP}
\label{sec:2-3-1}

\begin{figure}[t!]
\begin{center}
\includegraphics[width=0.5\textwidth]{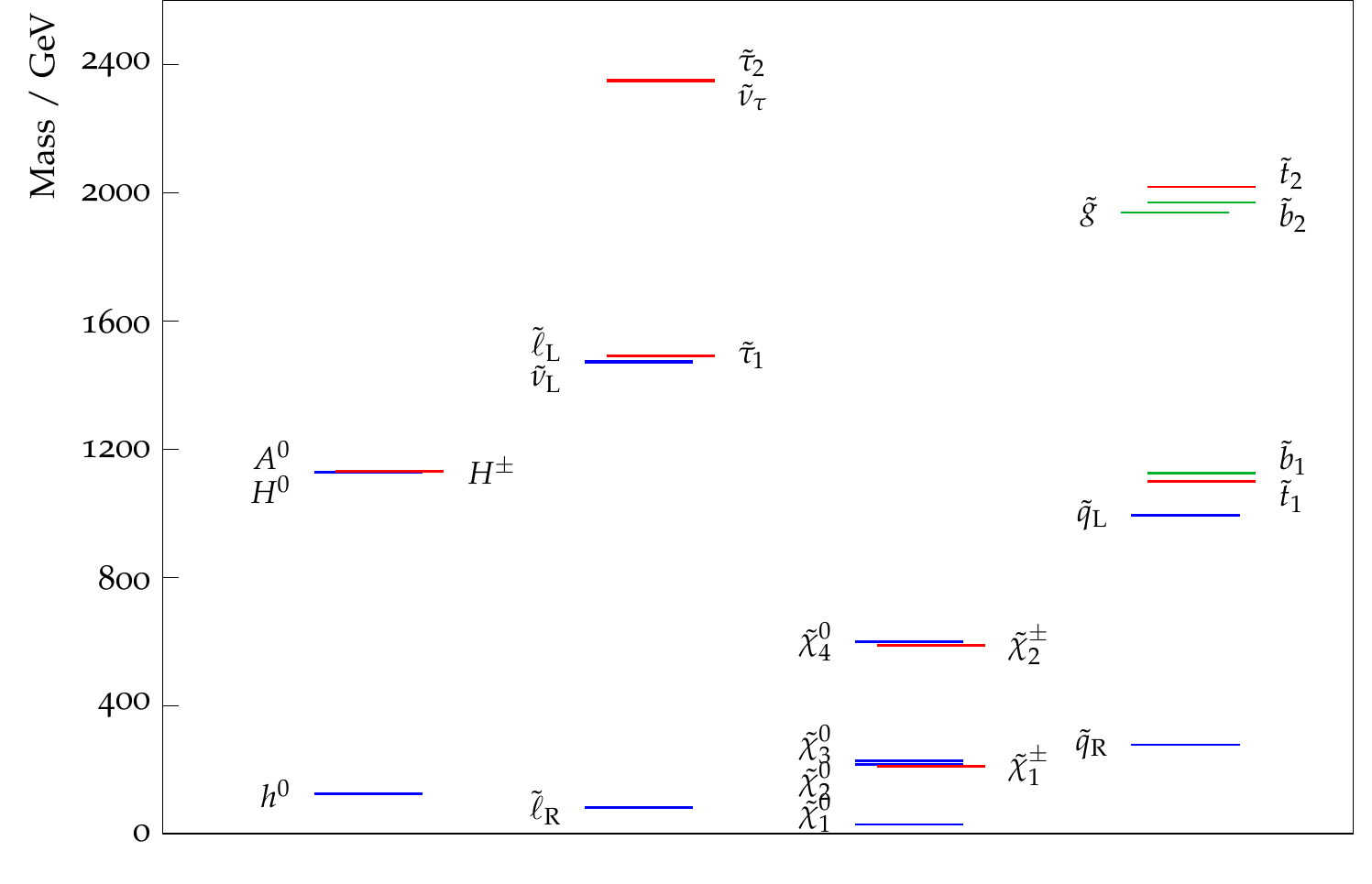}
\end{center}
\caption{Typical mass spectrum corresponding to the slepton NLSP scenario (class i) with $M_{\tilde \chi^0_1}$ = 
29.2~GeV and $M_{\tilde e_R,~\tilde \mu_R}$ = 80.1~GeV}
\label{fig:lowlspec}
\end{figure}

We consider here the case of points with slepton NSLP. An example of a viable mass spectrum is given in Fig.~\ref{fig:lowlspec}. 
Since the mass limits from LEP-2 are higher for left-handed sleptons, right-handed sleptons are favoured by the relic density constraint, 
which requires a small splitting. Beyond the small mass of the NLSP sleptons, this scenario remains relatively standard. However, a
neutralino mass of about at least 20~GeV is needed to accommodate the upper relic density bound, as can be seen from Fig.~\ref{fig:Oh2}, since 
the $\Delta M$ splitting remains relatively large.
Therefore, this scenario has only a limited interest corresponding to a region which is inside the exclusion contours by the CDMS 
and XENON experiments.

We also check the cosmological constraints by considering the indirect detection constraints by Fermi-LAT~\cite{Ackermann:2011wa}. The points passing all the constraints have neutralino annihilation cross-sections times relative velocity to $q\bar{q}$ smaller than $2\times10^{-30}$ cm$^3$/s, which is several orders of magnitude below the current Fermi-LAT limits and makes them compatible also with dark matter indirect detection limits.

\subsubsection{$\tilde \chi^{\pm}_1$ NLSP}
\label{sec:2-3-2}

Next we consider points where the NSLP is the lightest chargino, $\chi^{\pm}_1$ and the $\Delta M$ 
mass splitting is small.
This scenario corresponds generally to a compressed gaugino spectrum, with a small splitting between the chargino 
and neutralino masses. To accommodate the upper bound of the relic density constraint, a mass splitting 
of a few GeV between the lightest chargino and the lightest neutralino is generally required. 
As a consequence, the spin independent $\chi$-$p$ scattering cross-section is predicted to be 
relatively small, of order $10^{-6}-10^{-7}$~pb.
Since this scenario has a small spin independent $\chi$-$p$ scattering cross-section, it is marginal in 
accommodating the claims for WIMP direct detection and corresponds to a region well inside the CDMS and 
XENON exclusion curves. Furthermore, the production cross-section of $\chi^+_1 \chi^-_1$ and $\chi^0_2 \chi^0_1$ 
in this scenario at LEP-2 is large, 2.5 to 16~pb, and the detection efficiency of the LEP-2 analyses is 
$\sim$ 0.015 - 0.035, corresponding to about 20 to 250 detected signal events. The cross-section upper limit 
from the combination of the data of the LEP-2 experiments \cite{ADLO:2002aa} in this mass region is $\le$1~pb 
excluding all the points selected for this scenario.

\subsubsection{Light $\tilde q_{L,R}$ NLSP}
\label{sec:2-3-3}

\begin{figure}[t!]
\begin{center}
\includegraphics[width=0.5\textwidth]{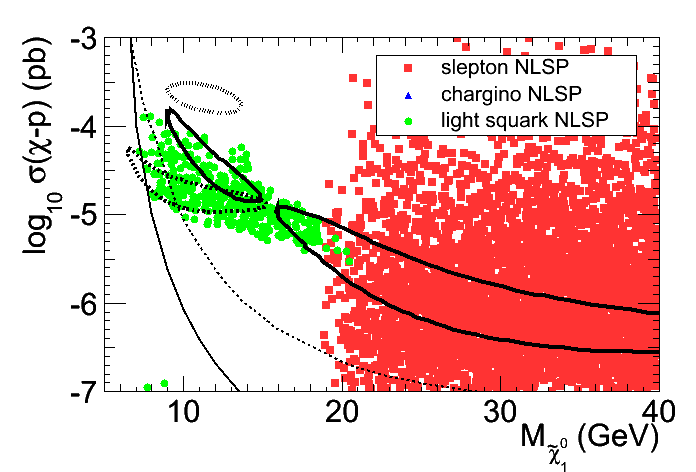}
\end{center}
\caption{Spin independent $\chi$-p scattering cross-section as a function of the $\tilde \chi^0_1$ mass. The points presented here pass all 
the previous constraints, including the tight relic density bounds. The red squares correspond to a slepton NLSP with a mass slightly above the 
LEP limits (class i), the blue triangles to scenarios with a chargino NLSP (class ii), and the green points to 
cases where a scalar quark is degenerate with the light neutralino (class iii).}
\label{fig:sig_classes}
\end{figure}

We move over to the case of a scalar quark NLSP with small mass splitting with the lightest neutralino.
The upper bound on the relic density constraint imposes a $\Delta M$ value of a few GeV. 
The degenerate squark can be any of the squarks with the exception of the scalar top, since this is required to be heavier 
to accommodate the $h^0$ mass range highlighted by the LHC data.
This scenario is the most interesting as it can provide us with a large spin independent $\chi$-$p$ scattering cross-section 
associated to a small neutralino mass (see Fig.~\ref{fig:sig_classes}). However, the couplings of the light squark to the 
$Z$ and $h^0$ bosons are in general large. In particular, the $Z$ decay width into squarks excludes this scenario, unless 
the squark decouples from the $Z$. This happens for specific values of the squark mixing angle. Since the first and second 
generation squarks do not mix, they are excluded leaving only a degenerate scalar bottom $\tilde b_1$ as a viable scenario.

\begin{figure}[t!]
\begin{center}
\includegraphics[width=0.39\textwidth]{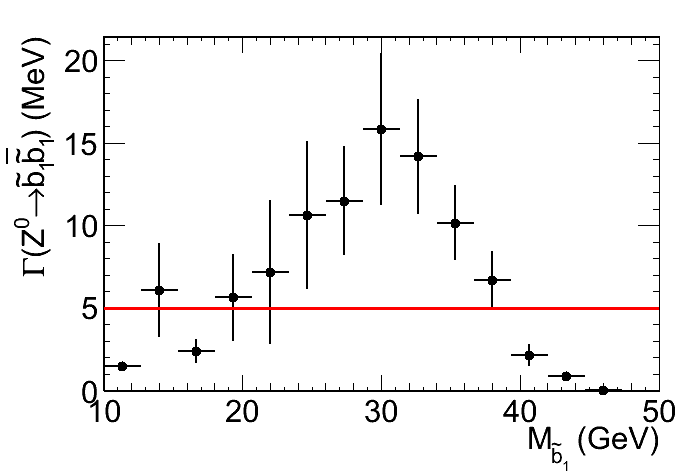}\\
\includegraphics[width=0.39\textwidth]{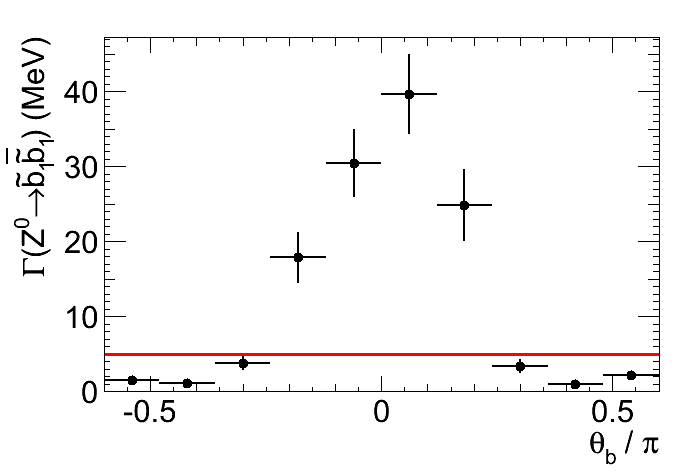}
\end{center}
\caption{Average $Z$ decay width to $\tilde b_1 \bar{\tilde b}_1$ in function of the $\tilde b_1$ mass (upper panel) 
and the sbottom mixing angle $\theta_b$ (lower panel). The horizontal lines correspond to the experimental limit.}
\label{fig:Ztheta}
\end{figure}
\begin{figure}[t!]
\begin{center}
\includegraphics[width=0.5\textwidth]{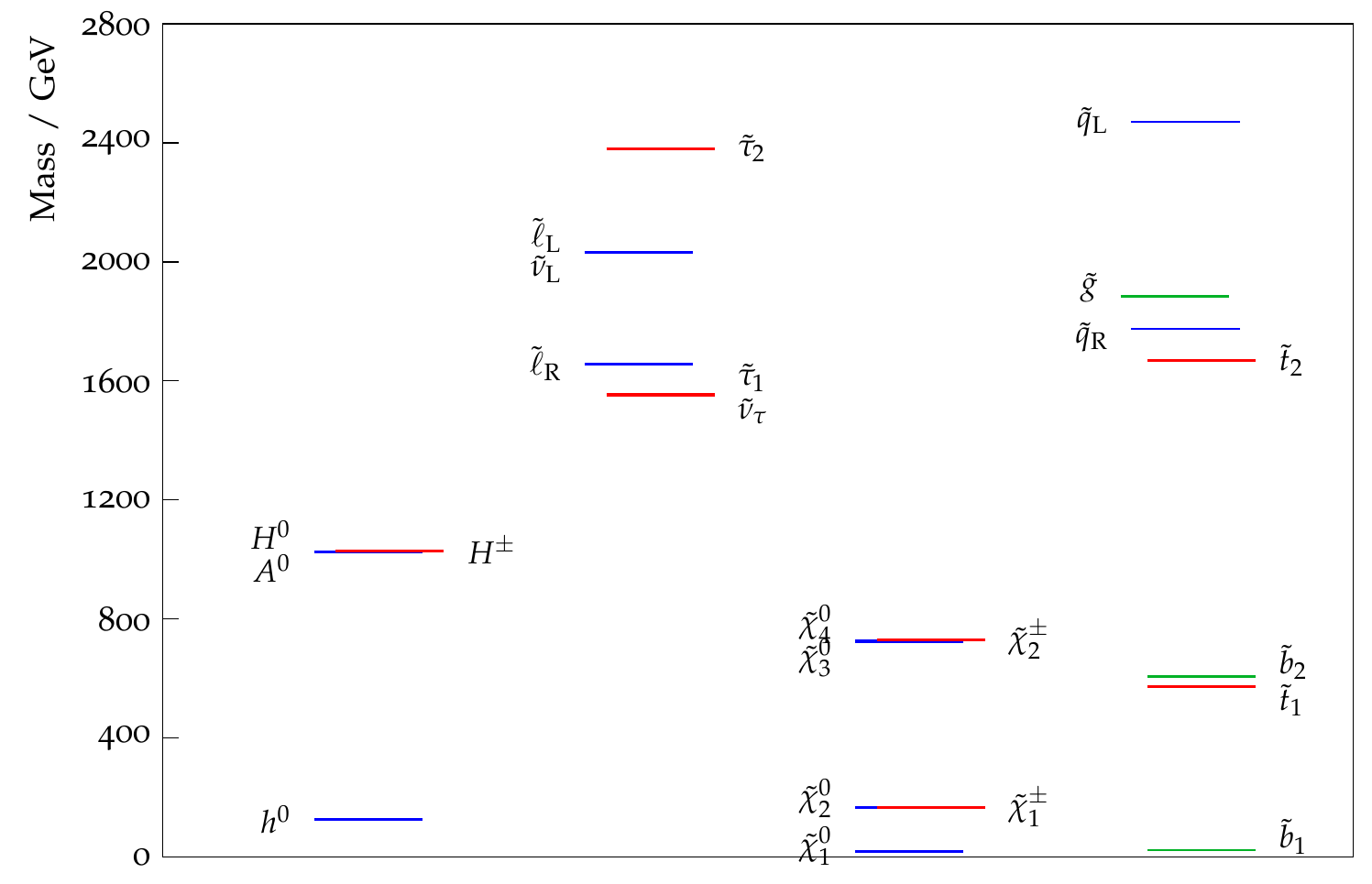}
\end{center}
\caption{Typical mass spectrum corresponding to the light sbottom degenerate with the lightest neutralino (class iii) 
with $M_{\tilde \chi^0_1}$ =18.1~GeV and $M_{\tilde b_1}$ = 22.9~GeV}
\label{fig:lowbspec}
\end{figure}

The third generation left-handed squarks have a common mass in the pMSSM and a very light $\tilde b_1$ is only possible if 
the right-handed bottom squark is very light. In this case, the mixing angle $\theta_b$ is large, close to $\pi/2$ corresponding 
to a mainly right-handed $\tilde b_1$, and the squark naturally decouples from the $Z$. The concurrence of a low value of the 
$\tilde b_1$ mass and its decoupling from the $Z$, through the mixing angle, is shown in Fig.~\ref{fig:Ztheta}. In addition, 
we observe that higher order SUSY corrections further decrease the $\tilde b_1$ mass, for appropriate values of the other parameters, producing typical spectra as that shown in Fig.~\ref{fig:lowbspec}. This ``sbottom miracle'' makes possible to 
find pMSSM solutions which reconcile a light neutralino signal at direct detection experiments with the LEP-1 constraints. 
At LEP-2 the pairs production of these light sbottom pairs has a cross-section of $\simeq$0.2~pb and a selection efficiency 
of 0.15 -- 0.40. The LEP-2 data \cite{ADLO:2004aa}  excludes sbottom pair production with $\tilde b \to b \tilde{\chi}^0$ at cross-section values above 
0.1~pb in this mass region and therefore rejects points in this scenario unless the $\Delta M$ mass splitting between the 
$\tilde b_1$ and the $\tilde\chi^0_1$ is smaller, or of the order of, the $b$ quark mass, where the selection efficiency of 
the LEP-2 analyses drops. In most of cases, the decay $\tilde b_1 \to \tilde\chi^0_1\,b$ is kinematically forbidden. 
At tree level, the most important open channel is $\tilde b_1 \to \tilde\chi^0_1\ s$, which is CKM-suppressed and may increase 
the $\tilde b_1$ lifetime up to a value comparable to that of $b$ hadrons. Furthermore, large QCD corrections can be expected 
for such a light state, and decay channels with larger decay rate can open up at higher orders, thus decreasing the lifetime.

\begin{figure}[t!]
\begin{center}
\includegraphics[width=0.5\textwidth]{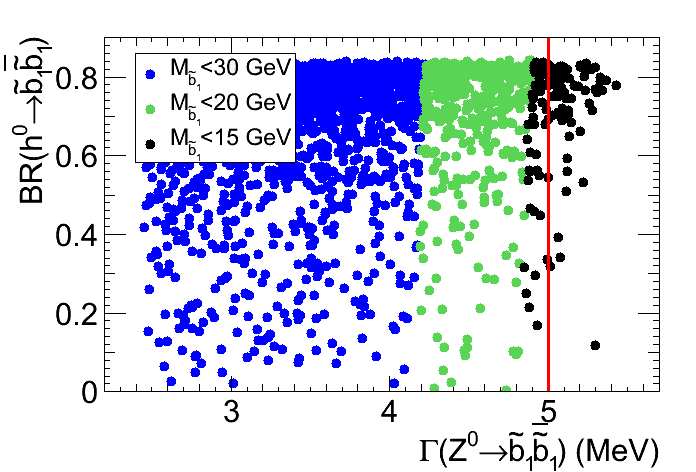}
\end{center}
\caption{$h^0$ branching fraction to $\tilde b_1 \bar{\tilde b}_1$
as a function of the $Z$ decay width to $\tilde b_1 \bar{\tilde b}_1$ for different values of the lightest sbottom mass.}
\label{fig:Zhiggs}
\end{figure}

Considering now this light sbottom and the Higgs boson, we notice that the rate of the decay 
$h^0 \to \tilde b_1 \bar{\tilde b}_1$ can become important in such scenarios. However, it is possible to find points for 
which the branching fraction of $h^0 \to \tilde b_1 \bar{\tilde b}_1$ and the $Z$ decay width to $\tilde b_1 \bar{\tilde b}_1$ 
are simultaneously small. In Fig.~\ref{fig:Zhiggs}, we show the correlations between the $Z$ decay width and the $h^0$ 
branching ratio to two photons. We remark that there is no strong correlations between both decays, and it is possible to have simultaneously a reduced $Z$ decay width to $\tilde b_1 \bar{\tilde b}_1$ and very small $h^0$ branching fraction to $\tilde b_1 \bar{\tilde b}_1$, even for $\tilde b_1$ masses as small as 15~GeV. Moreover, in this scenario, since the neutralino is mainly bino, the $h^0$ decaying to two light neutralinos is completely suppressed, resulting in an SM-like $h^0$ decay.

Finally, we check the cosmological constraints by considering the indirect detection constraints by 
Fermi-LAT \cite{Ackermann:2011wa}. The selected points corresponding to the degenerate $\tilde b_1$ scenario have neutralino 
annihilation cross-sections times relative velocity to $b\bar{b}$ smaller than $5\times10^{-27}$ cm$^3$/s, which is one order of magnitude below the 
current Fermi-LAT limits, which makes them compatible also with dark matter indirect detection limits.

In summary, after considering the constraint from the LEP data, the only viable scenario with a neutralino mass below 20 GeV corresponds to the light sbottom NLSP case.

In Fig.~\ref{fig:sig_classes}, we present distribution of the points passing the tight relic density bound.
\begin{figure}[t!]
\begin{center}
\includegraphics[width=0.5\textwidth]{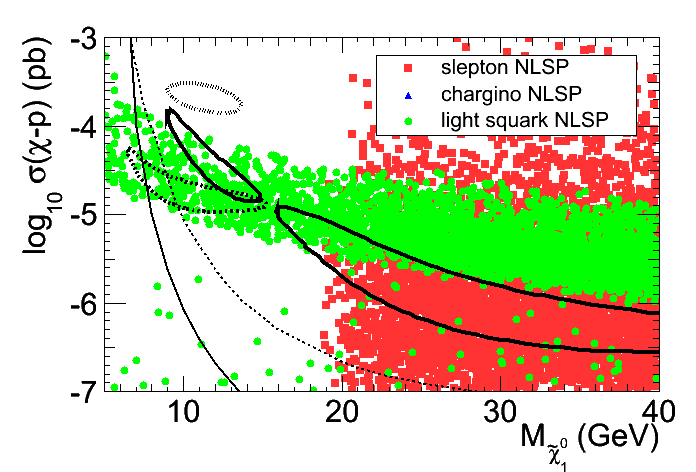}
\end{center}
\caption{Spin independent $\chi$-p scattering cross-section as a function of the $\tilde \chi^0_1$ mass. The points presented here pass all 
the previous constraints, including the loose relic density bound. The red squares correspond to a slepton NLSP with a mass slightly above the 
LEP limits (class i), the blue triangles to scenarios with a chargino NLSP (class ii), and the green points to 
cases where a scalar quark is degenerate with the light neutralino (class iii).}
\label{fig:sig_classes_upperomega}
\end{figure}%
Alternatively, in Fig.~\ref{fig:sig_classes_upperomega}, the same distribution is presented in the case where the loose relic density constraint is used.

A comparison of these two figures reveals that the lower bound of the relic density reduces the overall statistics, but also removes points corresponding to scenarios with a scalar quark degenerate with the light neutralino for neutralino masses above 20 GeV. This can be explained by the fact that points with a very small relic density have a small splitting. However, to get a relic density in the WMAP interval, the splitting should not be too small relatively to the neutralino mass. Also, the direct search bounds disfavour large splittings. Therefore, combining the relic density and direct search limits, only a small window remains where points can pass all the constraints.

\subsection{Non-standard scenarios}
\label{sec:2-4}

The calculation of the relic density and the dark matter direct detection constraints rely on many assumptions.
In particular, different cosmological scenarios can lead to a relic density which is larger than that computed 
in the standard cosmological scenario. First, the neutralino could be only one of several dark matter components.
Then, if dark energy were the dominant component at the time of the relic freeze-out, it  would result in an 
acceleration of the expansion of the Universe, which would lead to an earlier freeze-out and a much larger relic 
density~\cite{Kamionkowski:1990ni,Salati:2002md,Profumo:2003hq,Chung:2007cn,Arbey:2008kv}. Finally, entropy generation at the time of freeze-out, for example due to the decay 
of a late inflaton, can also lead to an increase -- or a decrease -- of the relic density~\cite{Moroi:1999zb,Giudice:2000ex,Fornengo:2002db,Gelmini:2006pq,Arbey:2009gt}. These effects are however limited by Big-Bang nucleosynthesis constraints, but using AlterBBN \cite{Arbey:2011nf}, we verified that they can nevertheless lead to an increase of three orders of magnitudes or more of the relic density while still being compatible with BBN constraints. 

Similarly to the relic density constraint, the direct detection constraints also rely on several assumptions. In particular, 
if the neutralino is not the only component of dark matter, or if the local density or velocity of dark matter is widely 
different from the standard assumptions\footnote{See for example \cite{Bidin:2012vt} for a discussion about the local 
density of dark matter.}, the constraints in terms of scattering cross-sections can be drastically changed. Therefore, 
even if the detection of a WIMP particles by a detector would permit to fix its mass, its scattering cross-section with matter 
would be dependent on large astrophysical uncertainties. The monojet searches at colliders however set constraints on the 
scattering cross-sections without suffering from astrophysical uncertainties.

\begin{figure}[t!]
\begin{center}
\includegraphics[width=0.5\textwidth]{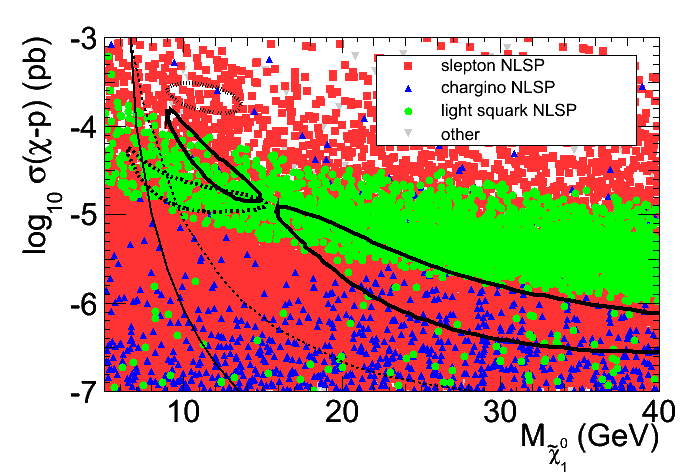}
\end{center}
\caption{Spin independent $\chi$-p scattering cross-section in function of the $\tilde \chi^0_1$ mass. The points presented here pass all the previous constraints, without applying the relic density limits. The red points correspond to a slepton NLSP with a mass slightly above the LEP limits (class i), the blue points to scenarios with a chargino NLSP (class ii), the green points to cases where a scalar quark is degenerate with the light neutralino (class iii), and the gray point to other more canonical scenarios.}
\label{fig:sig_classes_norelic}
\end{figure}

With these considerations, it is worth considering the results that we obtain by relaxing the relic density constraint. 
Fig.~\ref{fig:sig_classes_norelic} shows the different classes of points which pass all the constraints, without applying the relic density limits.
The three classes described above are well represented. In particular, the scenarios with degenerate squarks and sleptons are now 
realised over a broader range of scattering cross-sections and light neutralino mass values. In addition, the chargino NLSP scenario is allowed in this case. In particular, the relic density 
constraint imposes 
a small splitting between the neutralino and the NLSP. With this constraint removed, the splitting condition is strongly weakened, and it becomes possible to find compressed gaugino scenarios which pass LEP-2 $\chi^+_1 \chi^-_1$ and $\chi^0_2 \, \chi^0_1$ production constraint. Therefore these two classes of spectra can be rehabilitated if the relic density constraint is relaxed.

We also notice that other points which do not belong to any of the three classes can fulfill all the conditions. These scenarios have no other peculiarity than having a light neutralino and a scalar particle with a mass of a few hundreds of GeV, which can increase the scattering cross-sections.

\section{Sensitivity at LHC}
\label{sec:3}

In general terms, MSSM scenarios with a light neutralino offer no specific challenges 
to the LHC searches. Limits for $\tilde g$ and $\tilde q$ masses are commonly reported 
in the $M_{\chi^0_1}$ = 0 limit.  What makes the most viable scenario identified in this 
study specific for their search at the LHC is the high level of mass degeneracy between the 
LSP  $\tilde \chi^0_1$ and the $\tilde b_1$ scalar quark. This implies a very large production 
cross-section, of order 0.6~$\mu$b at 8~TeV accompanied by events with small transverse energy.

\begin{figure}[t!]
\begin{center}
\begin{tabular}{cc}
\includegraphics[width=0.22\textwidth]{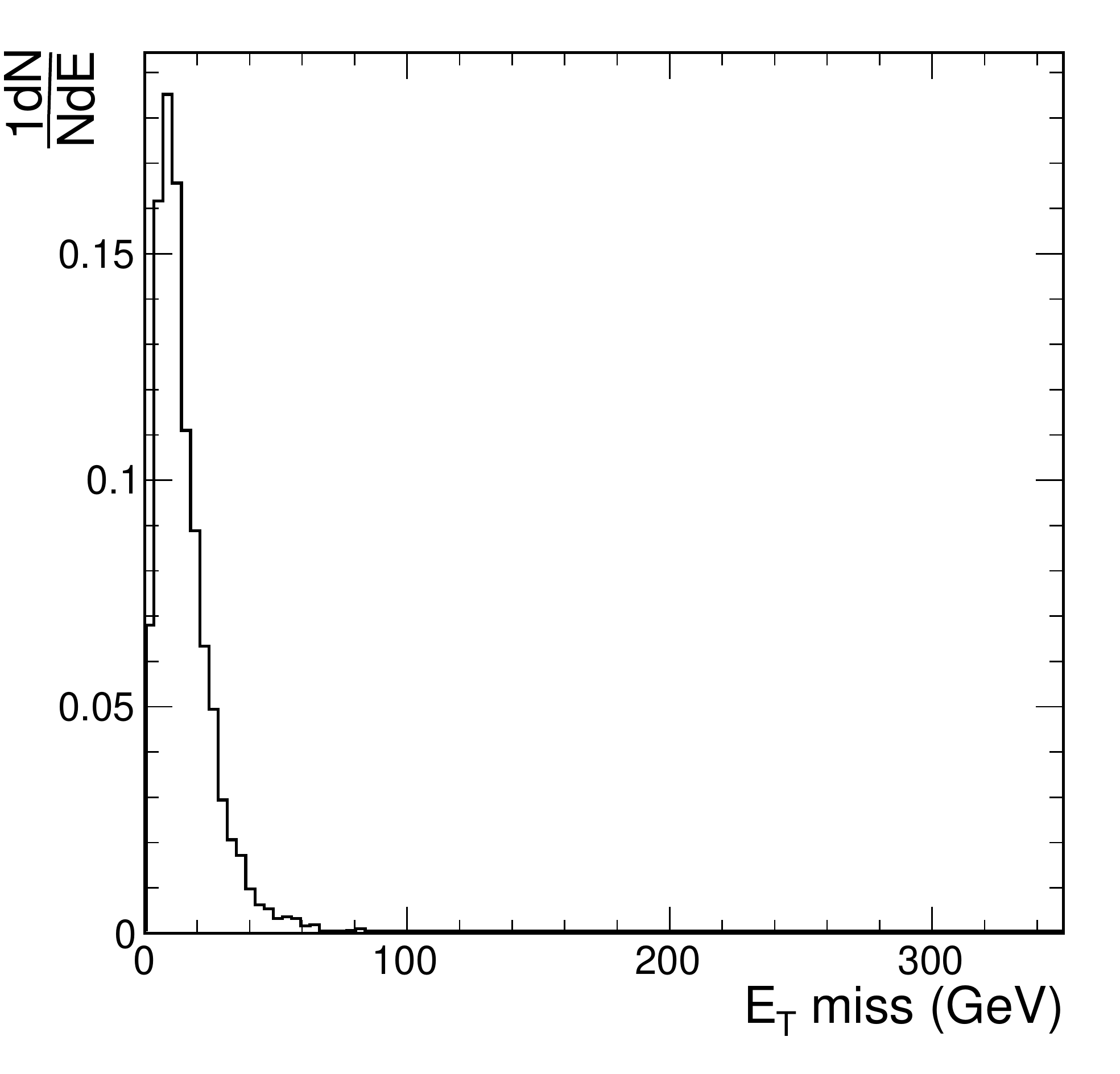} &
\includegraphics[width=0.22\textwidth]{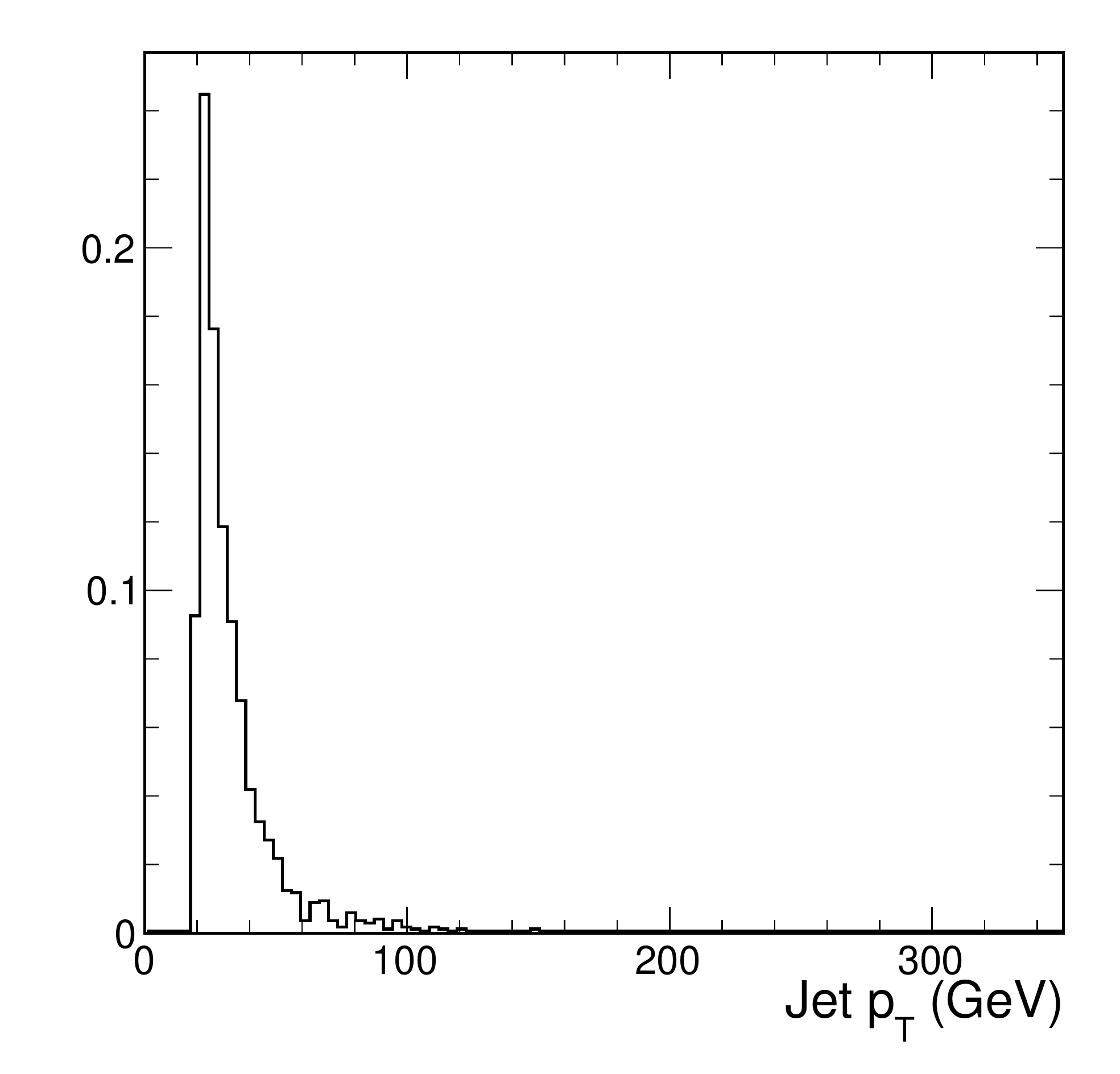} \\
\end{tabular}
\end{center}
\caption{Distributions of event missing $E_T$ (left) and jet $p_T$ (right) for $\tilde b_1$ production  
in 8~TeV $pp$ collisions in the almost degenerate light $\tilde b_1$ NLSP scenario.}
\label{fig:lhc}
\end{figure}

We study the distribution of the observables employed in the MET SUSY searches for a few 
points belonging to the various scenarios identified above. Here a word of caution is in order, since 
this analysis is carried out using fast simulation for events with a remarkably different kinematics 
compared to those used for its validation. Results can be considered valid in broader, qualitative 
terms but not necessary quantitatively. Fig.~\ref{fig:lhc} shows the distributions we obtain for 
the event missing $E_T$ and the jet $p_T$ for one of the selected points in the almost degenerate 
light $\tilde b_1$ NLSP scenario, which are relevant for the trigger and selection cut applied in MET 
analyses. 
By applying the jet $p_T$ and event missing $E_T$ cuts adopted in the recent searches for scalar bottom 
quark pair production with the ATLAS detector~\cite{Aad:2011cw,ATLAS:2012ah}, from a sample of 10k no events 
are kept. Similar strong cuts on missing $E_T$ and jet $p_T$ were applied in an earlier CMS search~\cite{11-006}. 
However, if the cuts on the two variables would be lowered from 130~GeV to 75 and 50~GeV, the selection 
efficiency would become of the order of 0.003 and 0.012, respectively.

\section{Conclusions}
\label{sec:4}

While the LHC experiments are progressively testing the SUSY parameter space for 
possible signals of strongly-interacting supersymmetric particle partners and 
they report events compatible with a Higgs signal, ground-based dark matter direct 
detection experiments offer a complementary approach in the search for a new 
weakly-interacting particle, which can be identified with the SUSY lightest 
neutralino. Three independent experiments have reported possible signals, which 
may be interpreted as due to the interaction of relic light neutralinos exhibiting a 
large scattering cross section, in their detectors. In this paper we have investigated 
the compatibility of such signals with the generic minimal supersymmetric extension 
of the Standard Model, taking into account the constraints from the low energy, flavour, 
LEP and LHC data. Through high statistics flat scans of the pMSSM we have identified 
several scenarios which give rise to very light neutralinos with large scattering 
cross section.  Once the LEP and LEP-2 limits are taken into account, the only viable 
scenario has the lightest scalar bottom quark, $\tilde b_1$ almost degenerate with the 
neutralino. This is compatible with the dark matter relic density from WMAP and indirect 
constraints from Fermi-LAT. A sizeable fraction of these points correspond to interaction 
cross section values close to the present constraints from CDMS and XENON. 
Relaxing the constraints from WMAP on the other hand leaves room also for other classes of scenarios 
such as the chargino or slepton NLSP cases to manifest themselves.

\section*{Acknowledgements}

We would like to thank M.~Mangano for supporting this activity and the LPCC for making dedicated  computing resources available to us. We are thankful to Ka Ki Li for her contribution during the early stage of this study. We acknowledge discussions with B.~Allanach, A.~Djouadi, A.~De Roeck,  G.~B\'elanger, E.~Nezri, G.~Polesello, K.~Rolbiecki, T.~Riemann, M.~Spira, E.~Aprile, P.~Beltrame, A.~Melgarejo and D.~Speller. We are also grateful to E.~Gianolio and the CERN IT Department for computing support.

\bibliographystyle{epjc}
\bibliography{susyscans2}
\end{document}